# SYNTHETIC CT GENERATION FROM TIME-OF-FLIGHT NON-ATTENUATION-CORRECTED PET FOR WHOLE-BODY PET ATTENUATION CORRECTION

*Weijie Chen, James Wang, Alan McMillan*

*University of Wisconsin-Madison*

## ABSTRACT

Positron Emission Tomography (PET) imaging requires accurate attenuation correction (AC) to account for photon loss due to tissue density variations. In PET/MR systems, computed tomography (CT), which offers a straightforward estimation of AC is not available. This study presents a deep learning approach to generate synthetic CT (sCT) images directly from Time-of-Flight (TOF) non-attenuation corrected (NAC) PET images, enhancing AC for PET/MR. We first evaluated models pre-trained on large-scale natural image datasets for a CT-to-CT reconstruction task, finding that the pre-trained model outperformed those trained solely on medical datasets. The pre-trained model was then fine-tuned using an institutional dataset of 35 TOF NAC PET and CT volume pairs, achieving the lowest mean absolute error (MAE) of 74.49 HU and highest peak signal-to-noise ratio (PSNR) of 28.66 dB within the body contour region. Visual assessments demonstrated improved reconstruction of both bone and soft tissue structures from TOF NAC PET images. This work highlights the effectiveness of using pre-trained deep learning models for medical image translation tasks. Future work will assess the impact of sCT on PET attenuation correction and explore additional neural network architectures and datasets to further enhance performance and practical applications in PET imaging.

## 1. INTRODUCTION

Positron Emission Tomography (PET) imaging relies on detecting photon pairs produced by positron-electron annihilation events in the body. These photons are emitted in opposite directions, and when they reach the PET detectors, they are recorded as "counts" that help reconstruct the spatial distribution of the radiotracer in the body. Time-of-Flight (TOF) PET imaging further enhances photon counting by measuring the time difference between the arrival of photon pairs at the detectors [1]. This additional timing information narrows down the probable location of the annihilation event along the line between the detectors, thereby improving spatial localization and reducing noise.

Attenuation correction (AC) is essential in PET imaging, as photon attenuation—caused by tissue density variations—can lead to signal loss, resulting in inaccurate tracer quantification. Traditional AC approaches often rely on computed tomography (CT) images to correct for photon attenuation. CT-based maps are effective because they provide highly detailed tissue density information, which translates to more accurate attenuation correction in PET imaging. In a PET-only or combined PET/MR scanner, CT is not available. However, techniques that generate a synthetic CT (sCT) generation from Non-Attenuation Corrected (NAC) PET images have been studied to create CT-equivalent attenuation maps without needing an actual CT scan, streamlining workflow and reducing radiation exposure [2, 3].

Deep learning has shown exceptional capabilities in tasks such as synthetic CT generation [4-6]. These methods typically leverage supervised image translation techniques to produce synthetic CT images that closely resemble true CT scans. However, obtaining a sufficiently large training dataset remains a challenge. Modern machine learning pipelines have emerged that involve training a foundational model on datasets containing millions of natural images, and then adapt the model to specific domains [7]. This approach provides a strong basis for capturing universal visual features and intrinsic relationships across images.

In this study, we employ the Vector Quantization (VQ) model [8], an architecture comprising an encoder, an embedding codebook, and a decoder. Input images are first processed through the encoder, which transforms them into a feature space. Vector quantization is then applied to discretize these continuous feature representations, storing only distinct codes in the codebook. The decoder subsequently reconstructs the images from these discrete codes, allowing the model to learn and retain unique patterns within the data effectively. This approach not only enhances the model's pattern recognition capabilities but also provides robustness to noisy inputs, making it well-suited for complex image reconstruction tasks.

This paper first compares the image reconstruction quality between models trained on a small, specific dataset and those pre-trained on a large natural image dataset, demonstrating the advantages of using pre-trained models. We then fine-tune the pre-trained model to perform TOF NAC PET to synthetic CT image translation, presenting both qualitative and quantitative results to highlight its effectiveness.

## 2. MATERIALS AND METHODS

### 2.1. Data acquisition

*2.1.1. Institutional dataset*

This study included 35 patient subjects, with data acquired on a Discovery 710 PET/CT scanner (GE Healthcare, Chicago, Illinois) for whole-body imaging. PET and CT data were collected simultaneously and registered, with all patient data anonymized prior to analysis. Raw PET data were reconstructed as time-of-flight (TOF) non-attenuation corrected (NAC) images using GE's Q.Clear reconstruction algorithm, a Bayesian penalized likelihood algorithm employing block sequential regularized expectation maximization (BSREM) for effective convergence [9]. A beta regularization parameter of 100 was applied to control the regularization strength. CT attenuation correction (CTAC) images were preprocessed to remove the table, allowing for a clean ground truth comparison in AI model development.

*2.1.2. Public datasets*

The two datasets used to pre-train the model are the smaller medical dataset, TotalSegmentator (TS), and the larger natural image dataset, OpenImages (OI). The dataset TotalSegmentator is collected for medical image segmentation, providing annotations for 104 anatomical structures in 1,204 CT scans, covering a diverse range of organs and tissues, with a total file size of 24 GB [10]. OpenImages, developed by Google, is a large-scale, versatile dataset containing approximately 9 million images that span a wide variety of real-world objects and scenes, with a total file size of 525 GB [11].

### 2.2. Model overview

The model trained on the TS dataset utilizes a pyramid VQ model [12]. To enhance performance, cosine similarity is applied to normalize embeddings on a unit sphere [13]. The embedding codebook is initialized using k-means clustering on the first batch of data. Additionally, a stale code expiration mechanism replaces unused embeddings with randomly selected vectors from the current batch, ensuring adaptability and codebook freshness [14].

The latent diffusion model (LDM) is trained as VQ models on the OI dataset [15]. We evaluate different LDM checkpoints and select the model with the lowest mean squared error (MSE) between input CT images and reconstructed CT images as the baseline for fine-tuning. Model configurations, including training from scratch, fully trainable, and encoder-frozen versions, are all tested to translate TOF NAC PET images to CT.

### 2.3. Implementation

*2.3.1. CT-to-CT reconstruction*

| Dataset | Model | Conv Depths | MAE of sCT in HU | | |
|---|---|---|---|---|---|
| | | | Soft | Bone | Whole |
| TS [10] | D3 | 3 | 42.52 | 158.23 | 54.05 |
| | D4 | 4 | 48.17 | 182.36 | 57.85 |
| OI [11] | f4 | 2 | 23.30 | 86.26 | 29.62 |
| | f4-noattn | 2 | 30.16 | 81.35 | 35.08 |
| | f8 | 3 | 35.87 | 134.04 | 46.69 |
| | f8-n256 | 3 | 67.58 | 268.77 | 76.00 |
| | f16 | 4 | 62.11 | 199.70 | 76.59 |

**Table 1.** MAE in HU for sCT reconstruction across different models and datasets (TS [10] and OI [11]), evaluated for soft tissue, bone, and whole image regions.

In the CT reconstruction stage, we evaluate the mean absolute error (MAE) between input and reconstructed CT images. The TS VQ model processes 64x64x64 voxel cubes from 1.5mm³ CT volumes in the TS dataset, using depths of 3/4 with starting convolutional channels set to 64, and the starting codebook sizes of 32. CT volumes are min-max normalized from [-1024, 2976] to [0, 1] and undergo random rotations and flips as data augmentation. Model performance is then compared on the institutional CT dataset.

Since the latent diffusion model (LDM) operates on 2D inputs, each 3D CT volume is sliced into 2D planes (axial, coronal, and sagittal), processed independently through the LDM model, and reconstructed into three separate volumes. A final reconstructed CT volume is generated by applying the median value across these three reconstructions for each voxel.

For precise evaluation, the MAE calculation is restricted to the body contour. This contour is obtained by first applying a mask to include only areas with Hounsfield Units (HU) over -500 across the entire volume. To fill any gaps within the body contour, the binary_fill_holes function from the SciPy package is applied slice by slice in the axial plane [16].

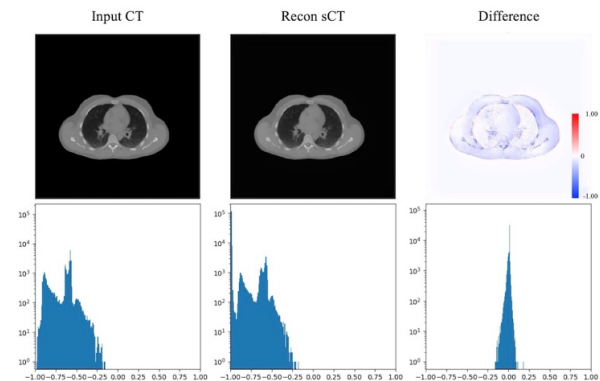

**Figure 1.** Visual results for the selected model (LDM-f4), showing the input CT, reconstructed synthetic CT (sCT), and the corresponding difference map in Blue-White-Red (BwR) color scale. The top row displays axial slices for each image, and bottom row shows the data distribution.

| Metrics | Scratch | | | No-frozen | | | Enc-frozen | | |
|---|---|---|---|---|---|---|---|---|---|
| | Whole | Soft | Bone | Whole | Soft | Bone | Whole | Soft | Bone |
| MAE ↓ | 92.449 | 62.415 | 279.364 | **82.204** | 56.196 | 287.812 | **74.488** | **53.826** | **228.396** |
| PSNR ↑ | 27.630 | 30.832 | 21.449 | 27.703 | **31.256** | 20.711 | **28.662** | **31.673** | **22.299** |
| SSIM ↑ | 0.867 | 0.912 | 0.618 | **0.872** | 0.918 | 0.613 | **0.880** | 0.919 | 0.620 |
| DSC ↑ | 0.978 | 0.940 | 0.527 | 0.979 | 0.943 | 0.506 | 0.981 | **0.950** | **0.630** |

**Table 2.** Quantitative performance comparison of three model configurations—Scratch, No-frozen, and Enc-frozen—based on MAE, PSNR, SSIM, and DSC across whole, soft tissue, and bone regions. Lower MAE and higher PSNR, SSIM, and DSC values indicate better performance. Statistically significant improvements over the Scratch model are in **bold**.

*2.3.1. PET to CT translation*

In this stage, we begin with the best LDM checkpoint and explore three model configurations using the same network architecture: training from scratch (Scratch), fully trainable (No-frozen), and encoder-frozen (Enc-frozen). Institutional PET-CT pairs are divided into five folds, with each trial using three folds for training, one for validation, and one for testing. PET and CT datasets are resampled to 1.5mm³ volumes, min-max normalized to a [-1, 1] range, and processed without data augmentation. The 3D PET/CT volumes are sliced in axial, coronal, and sagittal planes. We use L1 loss for training and optimize with the AdamW optimizer at a learning rate of 1e-5. All experiments are conducted on A100 GPUs. A Wilcoxon signed-rank test is conducted to compare No-frozen and Enc-frozen models to the Scratch model.

## 3. RESULTS

### 3.1. sCT Reconstruction

We compared models trained on the TS dataset (D3 and D4) with various LDM checkpoints (f4, f4-noattn, f8, f8-n256, f16), as shown in **Table 1**. The results indicate that the LDM-f4 model achieves the most accurate mapping for reconstructing input CT images. **Figure 1** presents visual results along with data distribution histograms, showing that the difference in pixel values is uniformly distributed around a mean of 0, suggesting no bias in the reconstruction. The pretrained encoder effectively maps images into a robust embedding space, while the decoder accurately reconstructs the embeddings back into images.

### 3.2. PET to CT translation

Three model configurations based on LDM-f4 checkpoints, Scratch, No-frozen, and Enc-frozen, are fine-tuned and evaluated using the metrics of MAE, Peak Signal-to-Noise Ratio (PSNR), Structural Similarity Index (SSIM), and Dice Similarity Coefficient (DSC), as shown in **Table 2**. Both the No-frozen and Enc-frozen models outperform the Scratch model, highlighting the benefits of pretraining. Visual results in **Figure 2** demonstrate improved synthesis of both the spine and soft tissue organs from PET images, particularly with the Enc-frozen model. The jagged boundaries are a result of fusing 2D slices into a 3D volume using the median method.

## 4. DISCUSSION

Previous work reported a non-masked MAE of 15.26 HU and a PSNR of 28.78 dB using a 3D UNet model with perceptual loss [5]. Another group achieved an MAE below 110 HU and a PSNR above 42 dB using CycleGAN [6]. Compared to these studies, the proposed workflow achieves the lowest MAE and highest PSNR when masked by the body contour.

This work has several limitations. Only one pre-trained framework, LDM, was used; exploring additional large datasets and neural network architectures could yield further improvements. Vision Transformers (ViT), which can handle larger datasets more effectively than convolutional neural networks, could be beneficial, as pre-trained ViT models for image generation may enhance performance. Additionally, fine-tuning the decoder may not fully leverage the pre-trained decoder's potential. Since reconstructed CT represents the upper limit for synthetic CT quality, developing an adapter to map embeddings from TOF NAC PET directly to CT—utilizing only the embeddings recorded in the VQ model's codebook—might further improve synthetic CT quality. Finally, applying this sCT as an attenuation map in downstream PET reconstruction would allow us to evaluate its practical impact on PET imaging accuracy.

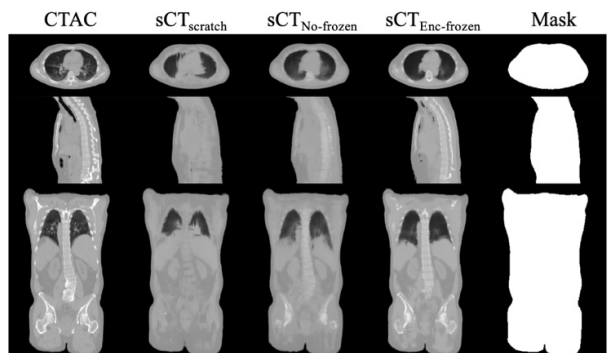

**Figure 2.** Example synthetic CT (sCT) images from the institutional dataset. From left to right: ground truth CT, sCT images from three different models, and the body contour mask used for computing metrics.

## 5. CONCLUSION

In conclusion, this project demonstrates the effectiveness of using a pre-trained model with fine-tuning configurations to generate high-quality sCT images from TOF NAC PET data. Through comparative analysis of model configurations, we show that leveraging pre-trained models significantly enhances sCT reconstruction accuracy, particularly when the encoder or full model is fine-tuned. The best-performing model configuration achieves notable improvements in MAE and PSNR when evaluated within the body contour, indicating its robustness and precision.

## 6. COMPLIANCE WITH ETHICAL STANDARDS

## 6. ACKNOWLEDGEMENT

This work was supported in part by NIH Grant R01EB026708 and the Department of Radiology at the University of Wisconsin. The Department of Radiology receives funding from GE HealthCare. Computational resources are partial provided by CHTC, UW-Madison [17]